\documentclass[letterpaper]{article}

\usepackage{cureport}
\usepackage{graphicx}
\usepackage{amsmath}
\usepackage{amssymb}
\usepackage{url}
\usepackage{multirow}
\usepackage{rotating}
\usepackage{subfigure}

\usepackage[toc,acronym]{glossaries}
\makeglossaries
\newacronym{snr}{SNR}{Signal to Noise Ratio}
\newacronym{eirp}{EIRP}{Effective Isotropic Radiated Power}
\newacronym{sinr}{SINR}{Signal to Interference and Noise Ratio}
\newacronym{los}{LOS}{Line of Sight}
\newacronym{nlos}{NLOS}{No(n) Line of Sight}
\newacronym{fcc}{FCC}{Federal Communications Commission} 
\newacronym{nist}{NIST}{National Institute of Standards and Technology} 
\newacronym{itm}{ITM}{Irregular Terrain Model}
\newacronym{itu}{ITU}{International Telecommunications Union}
\newacronym{itur}{ITU-R}{International Telecommunications Union Radiocommunication Sector}
\newacronym{rss}{RSS}{Received Signal Strength}
\newacronym{rssi}{RSSI}{Received Signal Strength Indicator}
\newacronym{ap}{AP}{Access Point}
\newacronym{ism}{ISM}{Industrial, Scientific and Medical}
\newacronym{unii}{U-NII}{Unlicensed National Information Infrastructure}
\newacronym{ssa}{SSA}{Spatial Simulated Annealing}
\newacronym{mle}{MLE}{Maximum Likelihood Estimator}
\newacronym{srs}{SRS}{Simple Random Sampling}
\newacronym{lcdb}{LCDB}{Landcover Classification Database}
\newacronym{dem}{DEM}{Digital Elevation Model}
\newacronym{dtm}{DTM}{Digital Terrain Model}
\newacronym{dsm}{DSM}{Digital Surface Model}
\newacronym{tia}{TIA}{Telecommunications Industry Association}
\newacronym{ntia}{NTIA}{National Telecommunications and Information Administration}
\newacronym{its}{ITS}{Institute for Telecommunications Sciences}
\newacronym{ccir}{CCIR}{International Radio Consultive Committee}
\newacronym{wmp}{WMP}{Wireless Measurement Project}
\newacronym{usgs}{USGS}{United States Geological Survey}
\newacronym{cost231}{COST-231}{European Cooperation in the field of Scientific and Technical Research Action 231}
\newacronym{sui}{SUI}{Stanford University Interim model}
\newacronym{rf}{RF}{Radio Frequency}
\newacronym{rfp}{RFP}{Request for Proposals}
\newacronym{cept}{CEPT}{European Conference of Postal and Telecommunications Administrations}
\newacronym{ecc}{ECC}{Electronic Communication Committee}
\newacronym{edam}{EDAM}{Effective Directivity Antenna Model}
\newacronym{pu}{PU}{Primary User}
\newacronym{vsa}{VSA}{Vector Signal Analyzer}
\newacronym{vsg}{VSG}{Vector Signal Generator}
\newacronym{fcs}{FCS}{Frame Check Sequence}
\newacronym{nda}{NDA}{Nondisclosure Agreement}
\newacronym{cpe}{CPE}{Client Premises Equipment}
\newacronym{poc}{POC}{Proof of Concept}
\newacronym{gps}{GPS}{Global Positioning System}
\newacronym{icmp}{ICMP}{Internet Control Message Protocol}
\newacronym{usb}{USB}{Universal Serial Bus}
\newacronym{http}{HTTP}{Hypertext Transport Protocol}
\newacronym{rmse}{RMSE}{Root Mean Square Error}
\newacronym{scrmse}{SC-RMSE}{Spread Corrected Root Mean Square Error}
\newacronym{ber}{BER}{Bit Error Rate}
\newacronym{per}{PER}{Packet Error Rate}
\newacronym{ofdm}{OFDM}{Orthogonal Frequency Division Multiplexing}
\newacronym{bps}{BPS}{Bits per Second}
\newacronym{cck}{CCK}{Complementary Code Keying}
\newacronym{agwn}{AGWN}{Additive Gaussian White Noise}
\newacronym{wand}{WAND}{Waikato Applied Network Dynamics}
\newacronym{wisp}{WISP}{Wireless Internet Service Provider}
\newacronym{isp}{ISP}{Internet Service Provider}
\newacronym{anova}{ANOVA}{Analysis of Variance}
\newacronym{aov}{AOV}{Analysis of Variance}
\newacronym{csma/ca}{CSMA/CA}{Carrier-Sense Multiple Access with Collision Avoidance}
\newacronym{isi}{ISI}{Inter-symbol Interference}
\newacronym{rms}{RMS}{Root Mean Square}
\newacronym{pdf}{PDF}{Probability Density Function}
\newacronym{cdf}{CDF}{Cumulative Distribution Function}
\newacronym{tfa}{TFA}{Technology For All}
\newacronym{wart}{WART}{Wide Area Radio Testbed}
\newacronym{cu}{CU}{University of Colorado}
\newacronym{bssid}{BSSID}{Basic Service Set Identifier}
\newacronym{bss}{BSS}{Basic Service Set}
\newacronym{bs}{BS}{Base Station}
\newacronym{oit}{OIT}{Office of Information and Technology}
\newacronym{siso}{SISO}{Single Input Single Output}
\newacronym{mimo}{MIMO}{Multiple Input Multiple Output}
\newacronym{csm}{CSM}{Channel State Matrix}
\newacronym{rem}{REM}{Radio Environment Map}
\newacronym{ls}{LS}{Least Squares}
\newacronym{wls}{WLS}{Weighted Least Squares}
\newacronym{aic}{AIC}{Akaike Information Criterion}
\newacronym{cr}{CR}{Cognitive Radio}
\newacronym{cmu}{CMU}{Carnegie Mellon University}
\newacronym{cw}{CW}{Continuous Wave}
\newacronym{gdal}{GDAL}{Geospatial Data Abstraction Library}
\newacronym{wgs84}{WGS84}{World Geodetic System 1984}
\newacronym{utm}{UTM}{Universal Transverse Mercator}
\newacronym{geni}{GENI}{Global Environment for Networking Innovation}
\newacronym{cat5}{CAT5}{Category 5}
\newacronym{visa}{VISA}{Virtual Instrument Software Architecture}
\newacronym{ni}{NI}{National Instruments}
\newacronym{cinr}{CINR}{Carrier to Interference and Noise Ratio}
\newacronym{evm}{EVM}{Error Vector Magnitude}
\newacronym{rce}{RCE}{Relative Constellation Error}
\newacronym{esnr}{ESNR}{Effective Signal to Noise Ratio}
\newacronym{pvc}{PVC}{Polyvinyl Chloride}
\newacronym{mbps}{Mbps}{Megabits per second}
\newacronym{idw}{IDW}{Inverse Distance Weighting}
\newacronym{mad}{MAD}{Mean Absolute Deviation}
\newacronym{psk}{PSK}{Phase Shift Keying}
\newacronym{qam}{QAM}{Quadrature Amplitude Modulation}

\TechReportYear{2011}
\TechReportMonth{September}
\TechReportNumber{1086-11}

\begin{document}

\title{The Stability of The Longley-Rice Irregular Terrain Model\\for Typical Problems}
\label{sec:bounding}

\author{Caleb Phillips, Douglas Sicker, and Dirk Grunwald}


\maketitle

\begin{abstract}
In this paper, we analyze the numerical stability of the popular Longley-Rice Irregular Terrain Model (ITM). This model is widely used
to plan wireless networks and in simulation-validated research and hence its stability is of fundamental importance to the correctness
of a large amount of work. We take a systematic approach by first porting the reference ITM implementation to a multiprecision framework and then generating loss predictions along many random paths using real terrain data. We find that the ITM is not unstable for common numerical precisions and practical prediction scenarios.
\end{abstract}




\section{Introduction}
\label{sec:itmstab}

The \gls{itm} is a well known and widely used model for predicting propagation loss
in long (greater than one kilometer) outdoor radio links. This model, in its most widely used incarnation, was developed by Hufford \textit{et al.} in \cite{ITMGuide} at the \gls{ntia} \gls{its} based on work done in 1978 by Longley \cite{Longley1978}. The model predicts the median attenuation of the radio signal as a function of distance and additional losses due to refractions at intermediate (terrain) obstacles. Compared to the vast majority of other models, even those that are similar in approach (e.g., The \gls{itu} Terrain Model \cite{Seybold2005}), the \gls{itm} is very complicated, requiring the interaction of dozens of functions that implement numerical approximations to theory. Despite its age, the \gls{itm} maintains a substantial popularity, possibly due to the fact that it was extensively validated in the 1970s for planning analog television transmissions, and that a reference implementation is available. Today, it is used in several popular coverage mapping and network planning tools (e.g., \cite{radiomobile,SPLAT}) as well as network simulation applications (e.g., \cite{qualnet,opnet}). As a result, it is also used widely in current wireless networking research (e.g., \cite{Harrison2010,Zennaro2010}), emerging government standards (e.g., \cite{FCC08260}), and persists at the foundation of other modern path loss models such as ITU-R 452 \cite{itur452}. Due to the ITM's complexity and its simultaneous popularity, the question of numerical stability is an obviously important one, but to our knowledge has not previously been investigated.

We take a systematic empirical approach to the analysis that involves porting the defacto C++ implementation
of the \gls{itm} \cite{ITMAlg} to a multiprecision framework. A comparison is made between the predicted path
loss values for many randomly generated links over real terrain data. Model parameters are also varied in order
to produce a fully factorial experimental design over a range of realistic parameters. In the end, the results show that while
the model performs disastrously for half-precision (16 bit) arithmetic, it is well behaved for single-precision (64 bit) 
and higher precisions. Within the values tested, there are only very few isolated cases that result in significantly 
different (greater than 3 dB) output and these tend to result from a single change in branching decision in the approximation algorithms 
and not because of massive information loss. While this sort if empirical analysis cannot be used 
to extrapolate to any parameters and any terrain model, we are able to say that over realistic links the model appears to 
be well behaved. This result provides confidence in the stability of the output of the \gls{itm} model as well as other similar
models that compute diffraction over terrain (e.g, \cite{Seybold2005,itur452}).

\section{Implementation}

The implementation involves a line by line port of the reference \gls{itm} implementation to have multiprecision support. By
and large, this involves using multiprecision data structures in place of native machine number formats. To this end, we make use of the
MPL, MPFR, and MPC libraries \cite{mpfr,gmp,mpc}. The MPFR library wraps the MPL library and provides additional
necessary features such as a square root function, computation of logs and powers, and trigonometric functions. The MPC library
provides support for complex arithmetic. In porting, a line like:

\begin{verbatim}
fhtv=0.05751*x-4.343*log(x);
\end{verbatim}

\noindent Must be replaced with:

\begin{verbatim}
mpfr_log(tmp,x,R);
mpfr_mul_d(tmp,tmp,4.343,R);
mpfr_mul_d(fhtv,x,0.05751,R);
mpfr_sub(fhtv,fhtv,tmp,R);
\end{verbatim}

\noindent Although it may be possible to automate this process, we proceeded manually in order to avoid introducing bugs. In porting, the refernce \gls{itm} source is modified to take an additional command line argument that specifies the precision in bits, which is passed to the multiprecision framework. Otherwise, the functionality and usage is identical to the machine-precision \gls{itm} implementation\footnote{Our multiprecision implementation is available at \url{http://systems.cs.colorado.edu/research/wireless/}.}. 

\subsection{Experiment}

\begin{figure}
\begin{center}
\includegraphics[width=0.7\columnwidth]{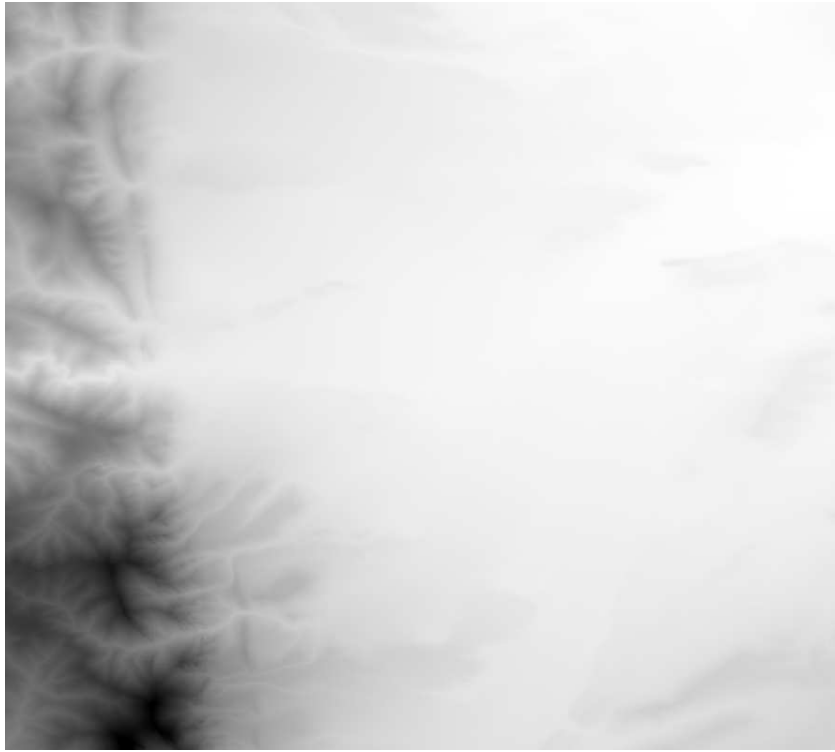}
\caption{Terrain (Digital Elevation Model) data for area within geographic bounding box. The plotted greyscale color varies from white (1562.15 m) to black (2550.28 m). Rolling hills and mesas (light grey) are at $\approx$ 1700 m, or 140m above the Colorado plateau.}
\label{fig:dem}
\end{center}
\end{figure}

\begin{table}
\begin{center}
\begin{tabular}{|l|l|}
\hline
Frequency (MHz) & 0.148, 80, 900, 1900, 2400, 5280, 60000\\
\hline
Climate Code & 1 (Equatorial), 2 (Continental Subtropical),\\
             & 3 (Maritime Subtropical), 4 (Desert),\\
             & 5 (Continental Temperate), 6 (Maritime\\
             & Temperate on land), 7 (Maritime\\
             & Temperate at Sea)\\
\hline
Permittivity/Conductivity & 5/0.001 (Poor Ground), 13/0.002, \\
                          & 15/0.005 (Average Ground), \\
                          & 25/0.02 (Good Ground), 80/5.0 \\
                          & (Sea Water) \\
\hline
Tx/Rx Height (Meters) & Uniform random: $U(0,35)$ \\
\hline
Tx/Rx Location & Uniform random within bounding box. \\
\hline
Precision (Bits) & 11, 24, 53, 64, 128, 256, 512, 1024, \\
                 & 32-bit Intel native \\
\hline
\end{tabular}
\caption{Experimental parameters\label{tab:ex}}
\end{center}
\end{table}

Our experimental design involves generating random link geometries within a latitude and longitude bounding box. 
For each random link, a path loss prediction is made both with the machine precision (64-bit double precision arithmetic) reference implementation and multiprecision implementation (at a variety of precisions). After the fact, we can quantify the differences 
in predictions and investigate any outliers or general trends.

For the bounding box we use 39.95324 to 40.07186 latitude and -105.31843 to -105.18602 longitude in the WGS84 datum.
This box contains a portion of the mountainous region to the west of Boulder, Colorado, as well as the plains to the east,
providing a realistic mix of topographies. 500 links are generated uniformly at random within the box. For each random link, we try a
range of reasonable parameter values, which are summarized in table \ref{tab:ex}. Antenna
heights are selected uniformly at random between 0 and 35 meters. For each link, the corresponding
elevation profile is extracted from a publicly available \gls{usgs} \gls{dem} with 0.3 arcsecond raster 
precision. Extraction and coordinate transformations are performed with the \gls{gdal} \cite{GDAL}. By trying each unique combination
of parameters with a random link and random transmitter heights, we must make 122,500 predictions at each precision. We evaluate 8 precisions ranging from half-precision (11 bits mantissa, 16 bits total) to 1024 bits along with the native 32-bit (64 bit double precision arithmetic) machine precision. This results in 1,102,500 predictions total total. Although this takes some time to compute, it can be parallelized trivially.

\begin{figure}[h!]
\begin{center}
\subfigure[All Precisions]{ \includegraphics[width=0.7\columnwidth]{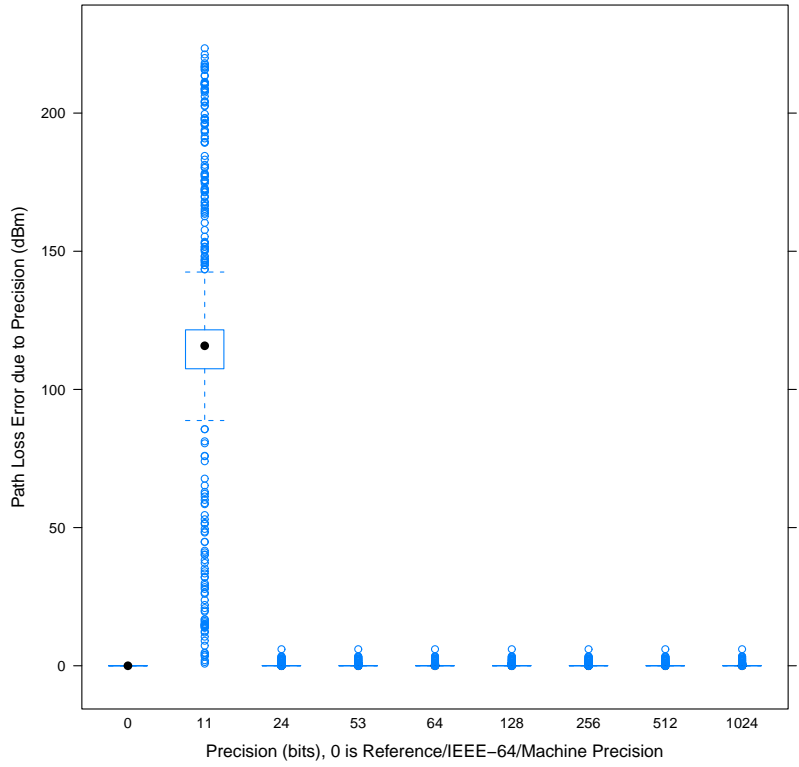}}

\subfigure[Single-Precision and Greater]{ \includegraphics[width=0.7\columnwidth]{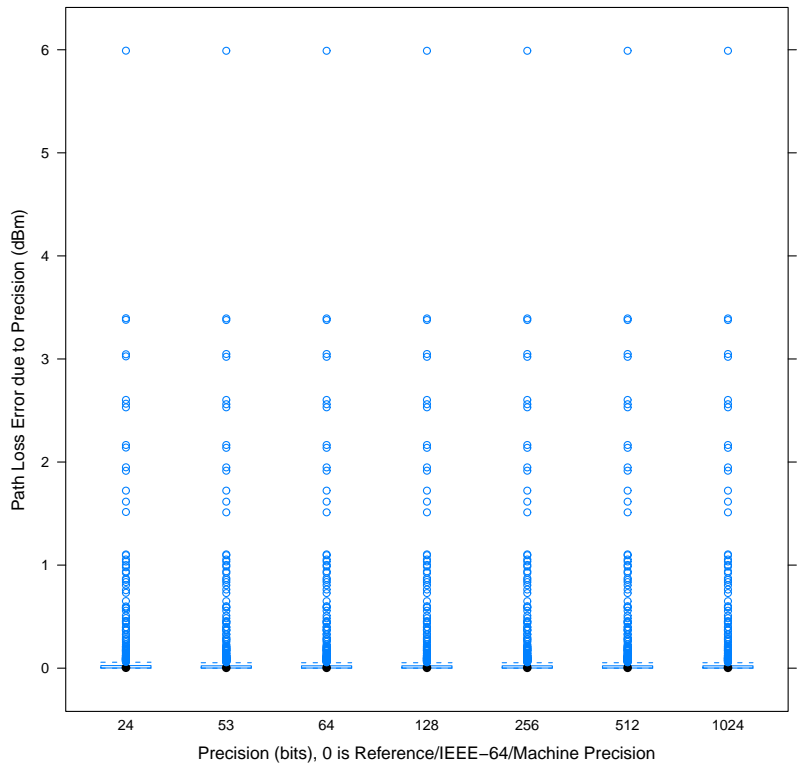}}

\caption{Box and whiskers plot of error as a function of precision. The zero-bit case is the machine-precision reference implementation.}
\label{fig:itmerr}
\end{center}
\vspace{-1em}
\end{figure}

\section{Results}

Figure \ref{fig:itmerr}a shows the overall results of this experiment: the error ($\epsilon$) between
the multiprecision prediction and the machine precision prediction is plotted. Half-precision arithmetic (11 bits of mantissa,
16 bits total) produces results that vary wildly. Above this, however, beginning at single precision (23 bits of mantissa, 1 bit of sign, 32 bits
total), the two programs make very similar predictions. Figure \ref{fig:itmerr}b provides a more detailed picture of these remaining cases. Much of the small error is negligible as it is presumably a function of differences in rounding\footnotemark. 
In the results, there is one clear outlier that produces a 6 dB difference. We investigated this case and found
that it was the result of a difference in a branching decision that chooses whether or not to make a particular correction. It is not clear
that one direction down the branch offers a better prediction than another, so this case can be safely ignored.

\footnotetext{IEEE 754-2008 requires subnormal arithmetic rounding, which is not done natively by the MPFR library. 
The majority of rounding (excluding this special case) are identical.}

Finally, we investigated the performance, in terms of running time, for the various precisions (on an otherwise unloaded machine). The multiprecision version is not substantially slower than the machine precision implementation---nearly all precisions take the same amount of time to run. At each precision, there are some number of outliers, which require slightly more time, but it is not clearly a function of precision. Hence, if it were the case that the multiple precision implementation was also safer, then its use would be clearly preferable.

\section{Conclusions}

Although it is not possible to extrapolate universally from these results, they demonstrate that the \gls{itm} is \textit{not} substantially
unstable for typical problems and reasonably precise numeric types (i.e., single and double precision IEEE formats). An analytical
investigation of stability would go a long way to determine the stability universally, but is a substantial undertaking that
involves the careful dissection of dozens of complex algorithms that combine to create the \gls{itm} implementation. A motivated investigator
may choose to focus his effort on the knife-edge diffraction approximation algorithm, which is almost certainly the most numerically
complex component of the model. For most practical purposes, however, the results presented here are sufficient to justify continued 
use of this model with the confidence that under typical situations it is not significantly affected by rounding and cancellation errors.

\clearpage
\bibliographystyle{IEEEtran}
\bibliography{proposal,me}



\end{document}